\documentclass[aps,prb,groupedaddress]{revtex4}

\usepackage{graphicx}
\usepackage{dcolumn}
\usepackage{epsfig}
\usepackage{bm}
\usepackage{amsfonts}
\usepackage{latexsym}
\usepackage[dvipdfm]{hyperref}

\newcommand{\be}{\begin{equation}}
\newcommand{\ee}{\end{equation}}
\newcommand{\bea}{\begin{eqnarray}}
\newcommand{\eea}{\end{eqnarray}}

\newcommand{\half}{ \frac{1}{2} }

\begin{document}


\title{Hard Core Bosons on the Triangular Lattice at Zero Temperature:\\
A Series Expansion Study}



\author{J. Oitmaa}
\email[]{j.oitmaa@unsw.edu.au}
\affiliation{School of Physics,
The University of New South Wales,
Sydney, NSW 2052, Australia.}

\author{Weihong Zheng}
\email[]{w.zheng@unsw.edu.au}
\homepage[]{http://www.phys.unsw.edu.au/~zwh}
\affiliation{School of Physics,
The University of New South Wales,
Sydney, NSW 2052, Australia.}

\author{D. Tompsett}
\affiliation{School of Physics,
The University of New South Wales,
Sydney, NSW 2052, Australia.}


%

\date{\today}

\begin{abstract}
We use high order linked cluster series to investigate the hard core boson model on the
triangular lattice, at zero temperature. Our expansions, in powers of the hopping parameter $t$,
probe the spatially ordered `solid' phase and the transition to a uniform superfluid phase.
At the commensurate fillings $n=\frac{1}{3}, \frac{2}{3}$ we locate a quantum phase transition point at
$(t/V)_c\simeq 0.208(1)$, in good agreement with recent Monte Carlo studies. At half-filling
($n=\frac{1}{2}$) we find evidence for a solid phase, which persists to $t/V\simeq 0.06$.
\end{abstract}




\maketitle

\section{\label{sec:intro}INTRODUCTION}

There is considerable interest at present in the physics of a system of hard-core bosons on the
two-dimensional triangular lattice, described by the Hamiltonian
\be
H = - t \sum_{\langle ij\rangle } ( a_i^\dagger a_j + a_j^\dagger a_i ) + V  \sum_{\langle ij\rangle } n_i n_j
- \mu \sum_i n_i \label{eq_H}
\ee
where each lattice site can be occupied by 0 or 1 bosons ($n_i=0,1$).
The first term represents hopping between nearest neighbour sites, the $V$-term is a nearest
neighbour repulsion, $n_i = a_i^\dagger a_i$, and $\mu $ is the chemical potential. This
Hamiltonian also arises as a limiting case ($U= \infty $) of the so-called `Bose-Hubbard' model,
where multiple occupancy is allowed, albeit at a cost in energy.

Interest is this model has arisen, in the main, because of the possibility of an exotic `supersolid'
phase where long-range crystalline order and superfluidity may co-exist. Such a phase
was conjectured over thirty year ago and has remained controversial and unobserved until,
perhaps, very recently\cite{kim04}. On the other hand evidence for supersolid phases has been found in lattice models of interacting
hard-core bosons, and the possibility of experimental realizations with
atoms in optical potentials\cite{gre02} provides an exciting prospect
of direct verification of the theoretical predictions.

As is well known, a lattice model of hard-core bosons with nearest-neighbour
repulsion can be mapped directly to a spin-$\half$ antiferromagnet,
in general in the presence of a magnetic field. Techniques developed and used for the
magnetic system can then be used to study the boson problem.
The early work of Liu and Fisher\cite{liu73} studied the phase diagram of the
model, on a bipartite lattice, using the usual mean-field approximation,
and showed the existence of a supersolid phase under certain conditions.
More recently Murty {\it et al.}\cite{mur97} have used first-order
spin-wave theory to study the model on the two-dimensional
triangular and kagome lattices, and also find
a stable supersolid phase. These lattices are attractive
candidates for exotic quantum phases because of frustration and low
dimensionality, and the consequent enhancement of quantum fluctuations.
Both of these studies\cite{liu73,mur97} involve, however,
approximations whose validity is difficult to judge.

In an attempt to avoid uncontrolled approximations Boninsegni\cite{bon03}
used a Green Function Monte Carlo method to study the triangular
lattice. He verified the presence of a supersolid phase, albeit within a smaller
range of parameters than predicted at the mean-field level.
In particular, he found for $V/t \gtrsim 7$, a narrow supersolid phase at and above
the commensurate filling $n=\frac{1}{3}$ ($n\equiv \langle n_i \rangle $
is the average boson number density), and by particle-hole symmetry, also at and below $n=\frac{2}{3}$.
He found no supersolid phase at $n=\frac{1}{2}$ for any value of $V$. More recent
work\cite{hei05,mel05,wes05,bon05} has further clarified the phase
diagram, particularly at and near half-filling ($n=\frac{1}{2}$). It
appears, contrary to earlier expectations, that the supersolid phase
persists at $n=\half$ although its spatial symmetry remains controversial.

Our goal in this paper is to investigate the ground state properties
of the triangular lattice hard-core boson model via
the technique of high-order series expansions\cite{gel00,oit06},
an approach complementary to the Monte Carlo work. A particular advantage of
the series  approach is the ability to locate critical points reliably
and accurately.
To the best of our knowledge the hard-core boson problem has
not previously been studied using this approach, although there has been some work on the
Bose-Hubbard model without nearest neighbour repulsion\cite{els99}.

The paper is organized as follows. Section \ref{sec2} discusses some
general aspects of the model and briefly describes the series expansion methodology. Section \ref{sec3} gives our results for the case of $n=\frac{1}{3}$
and $\frac{2}{3}$, and discusses these in relation to existing work.
Section \ref{sec4} considers the half-filling case $n=\half$.
Finally in Section \ref{sec5} we summarize our results and attempt to draw conclusions.

\section{\label{sec2}General Features and Methodology}
As already mentioned above, the hard-core boson Hamiltonian has a particle-hole
symmetry. Under the transformation
$\bar{a}_i=a_i^\dagger$, $\bar{n}_i = 1 - n_i$ the Hamiltonian (\ref{eq_H})
becomes
\be
\bar{H}= N (\half z V - \mu ) - t \sum_{\langle ij \rangle} (\bar{a}_i^\dagger
\bar{a}_j + \bar{a}_j^\dagger \bar{a}_i ) +
V \sum_{\langle ij \rangle} \bar{n}_i \bar{n}_j
- (zV -\mu) \sum_i \bar{n}_i
\ee
which, apart from a constant, is identical in form to the original. Here
$z(=6)$ is the coordination number of the lattice. Thus the phase diagram is symmetric
about the half-filled case $n=\half$.
Consequently our series for the commensurate filling $n=\frac{1}{3}$ are unchanged
for $n=\frac{2}{3}$, and we consider the two cases together in Section \ref{sec3}.

It is also useful to display the equivalent spin Hamiltonian, obtained via the transformation
\be
a_i \to S_i^-, \quad a_i^\dagger \to S_i^+, \quad n_i \to S_i^z + \half
\ee
which yields, apart from a constant
\be
H_{\rm spin} = V \sum_{\langle ij \rangle} S_i^z S_j^z
- 2 t \sum_{\langle ij \rangle} (S_i^x S_j^x + S_i^y S_j^y )
- (\mu - \half zV) \sum_i S_i^z \label{eq_Hspin}
\ee
i.e., a Heisenberg model with exchange anisotropy (XXZ model) in a magnetic field.
A phase with $\langle S^z \rangle \neq 0$, $\langle S^x \rangle=0$ then corresponds to
a solid phase or normal fluid, while a phase with non-zero magnetization
in the x-y plane corresponds to a superfluid (if $\langle S^z\rangle$ is uniform)
or a supersolid (if $\langle S^z \rangle$ is spatially modulated).

The series expansion method is based on writing the Hamiltonian in the form
\be
H = H_0 + \lambda H_1
\ee
where $H_0$ is exactly solvable and $H_1$ is treated perturbatively.
A linked-cluster approach\cite{gel00,oit06} is the most efficient,
permitting the series to be derived to quite high order.
Here we take the nearest neighbour repulsion term as $H_0$, and
the hopping term as the perturbation.  The Hamiltonian (\ref{eq_H})
is particle conserving and hence the chemical potential term can be dropped in the calculations.
For convenience we also set $V=1$ to fix the energy scale.
The perturbation parameter is then
$\lambda = -t$, and our series are obtained in powers of $t$.
As the series are of finite length we are unable to effectively probe
the large $t$ region of the phase diagram.
In magnetic language, our unperturbed state is a commensurate
N\'eel type state, and x-y spin fluctuations are included perturbatively.
We are unable to treat the superfluid/supersolid phases directly.
Nevertheless, as we shall show, a great deal of useful information
about the model can be obtained.

\begin{figure}[htb]
\centering
      \includegraphics[width=0.8\textwidth]{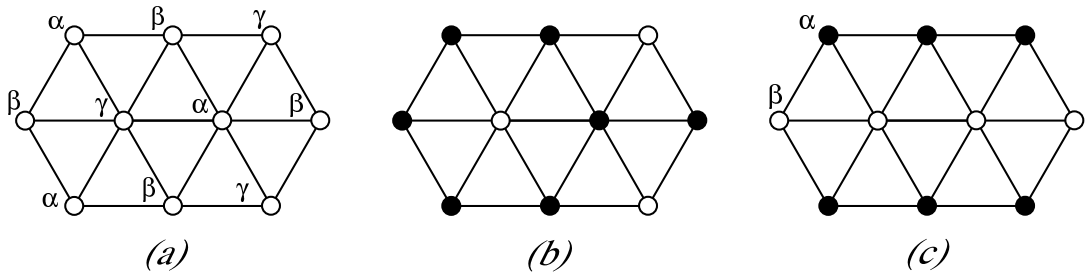}
      \caption{(a) The 3-sublattice decomposition of the triangular lattice;
      (b) Unperturbed configuration for $n=\frac{2}{3}$ (the filled circle represent
      occupied sites); (c) Unperturbed configuration for $n=\half$. \label{fig1}}
\end{figure}

The triangular lattice can be decomposed into three equivalent sublattices,
which we denote $\alpha$, $\beta$, $\gamma$, as shown in Fig. \ref{fig1}(a).
For filling $n=\frac{2}{3}$ the unperturbed ground state has two sublattices
(say $\alpha$ and $\beta$) fully occupied and the third ($\gamma$) empty. This corresponds to a solid phase, with a density
modulation with wavevector ${\bf k} = ({4 \pi\over 3}, 0)$, and is shown
in Fig. \ref{fig1}(b). It is also referred to as the
$\sqrt{3}\times \sqrt{3}$ phase. Although the ground state is threefold
degenerate, these do not mix in any finite order of perturbation theory and
we can start from one of the partners. Turning on the hopping term allows particles to occupy the other sublattice,
and at some critical value $t_c$ the average sublattice occupancies will
become equal. This represents the transition from solid to superfluid. An
order parameter can be defined as
\be
m \equiv  \langle n_{\alpha} \rangle - \langle n_{\gamma} \rangle = 3 \langle n_{\alpha}\rangle -2  \label{eq_m}
\ee
which equals 1(0) in the fully ordered (disordered) phase. The transition can
also be located from the behaviour of nearest neighbour correlators
$C_{\alpha\beta} \equiv \langle n_{\alpha} n_{\beta} \rangle $, or from the static
structure factor
\be
S({\bf k}) = {1\over N} \sum_{ i,j } \langle n_i n_j\rangle
 e^{i {\bf k} \cdot ( {\bf R}_j- {\bf R}_i ) } \label{eq_S}
\ee
at ${\bf k}={\bf k}^*$, where ${\bf k}^* = ({4\pi\over 3},0 )$ is the wavevector for the
$\sqrt{3}\times \sqrt{3}$ phase.
For $n=\frac{1}{3}$ a completely analogous treatment can be used, with the
unperturbed ground state having one sublattice (say
$\gamma$) fully occupied, and the other sublattices empty.


At half-filling ($n=\half$) the unperturbed system is equivalent to an isotropic
Ising antiferromagnet in zero field, which is known to be disordered
at $T=0$, with a macroscopic ground state degeneracy. It seems,
however, that a nonzero hopping term lifts this degeneracy and leads to a stable
supersolid phase - an example of the `order from disorder' phenomenon\cite{sheng92},
induced by quantum fluctuations. To study this case via series expansions
we include an additional field in the unperturbed Hamiltonian (and subtract it
via the perturbation term), to choose a unique ground state. We
choose the striped state shown in Fig. \ref{fig1}(c), for which a two-sublattice
decomposition of the triangular lattice is made.
This will be discussed further in Section \ref{sec4}.

\section{The $n=\frac{1}{3}, \frac{2}{3}$ phases \label{sec3}}
We turn now to our results, considering first the commensurate phase with $n=\frac{2}{3}$.
The unperturbed ground state ($\sqrt{3}\times \sqrt{3}$) is shown in Figure \ref{fig1}(b).
The order parameter $m$ (Eqn. \ref{eq_m}) will decrease with increasing $t$, and,
assuming a normal 2nd order transition,
will go to zero with a power law
\be
m \sim (t_c - t)^{\beta}, \quad t \to t_c^-
\ee
where $\beta$ is a critical exponent.

We have computed a series expansion for $m$, up to order $t^{12}$.
The series coefficients are given in Table 1. Dlog Pad\'e approximants (shown in Table \ref{tab_pade})
and integrated differential
approximants\cite{gut} have been used to analyse this series,
with the resulting estimates
\be
t_c=0.21(1),  \quad  \beta=0.09(1)    \label{eq_t_c}
\ee
The estimate $t_c=0.21$ is consistent with the most recent Monte Carlo estimate\cite{wes05} of
$0.195\pm 0.025$. We are unaware of any previous estimate of the exponent $\beta$.
The order and universality class of this transition remain unclear.
Our exponent estimate $\sim 0.09(1)$ is similar to that of the 3-state
Potts model in two dimensions ($\beta =1/9$). On the other hand the 3-state
Potts model in three dimensions is believed to have a weak first-order transitions.

It would be straightforward to compute the order parameter for specific values of $t$,
and to display a curve of $m(t)$, but we do
 not do this.

\begin{figure}[htb]
\centering
      \includegraphics[width=0.5\textwidth]{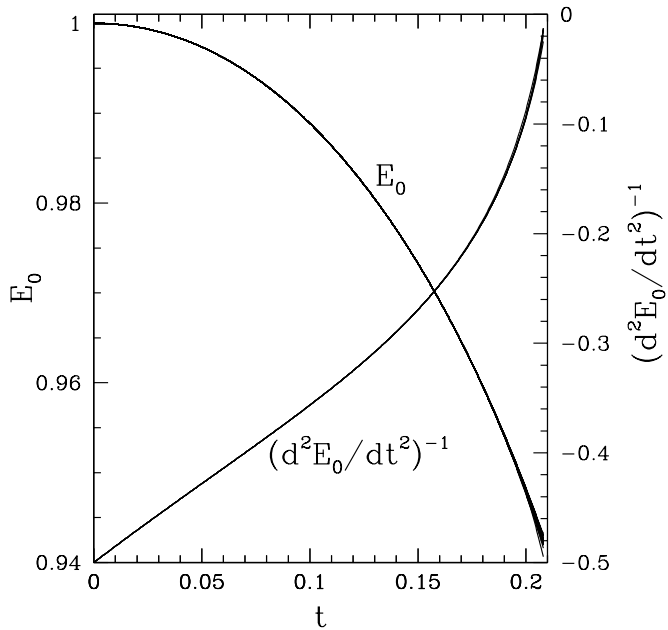}
      \caption{The ground state energy per site $E_0$ and $(d^2 E_0/dt^2)^{-1}$ versus $t$
       for $n=2/3$ filling, the results of
       several different integrated differential approximants to
            the series  are shown. \label{fig_e0_c}}
\end{figure}

\begin{figure}[htb]
\centering
      \includegraphics[width=0.5\textwidth]{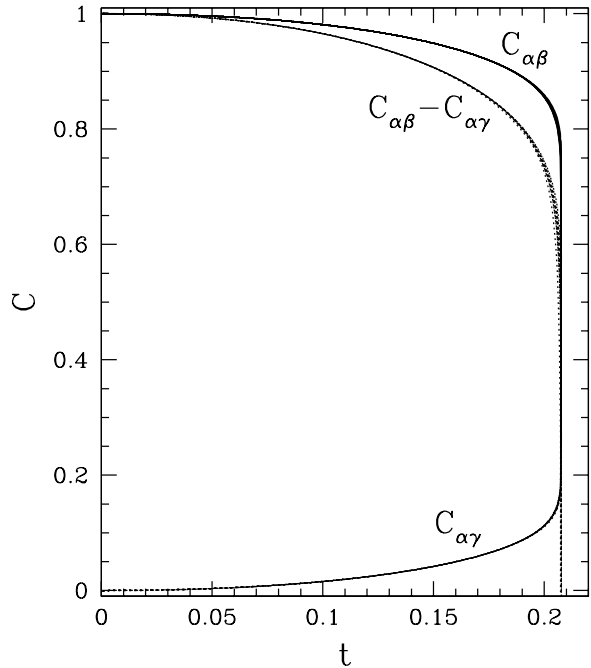}
      \caption{The nearest neighbor correlators  $C_{\alpha \beta}$ and
      $C_{\alpha \gamma}$ and their difference, versus $t$,
       for $n=2/3$ filling.  Several different integrated differential approximants to
            the series  are shown. \label{fig_C}}
\end{figure}

 An alternative way of estimating the critical point is to compute the series
 for the ground state energy $E_0 (t)$ and for the second derivative $d^2 E_0/dt^2$
 (analogous to a `specific heat'). This analysis is less precise, but the results are
 consistent with (\ref{eq_t_c}). In Figure \ref{fig_e0_c} we show $E_0$ and the inverse of the
 second derivative as functions of $t$, for $0<t\leq t_c$.
 The results of  several different integrated differential approximants to
            the series  are shown. The second derivative is seen  to diverge at
            $t\sim 0.21$.

 We have also obtained series in $t$ for the pair correlations $C_{ij} \equiv \langle n_i n_j \rangle$.
 One such series is given in Table \ref{tab_ser}) - we are happy to provide others on request.
Figure \ref{fig_C} shows three curves,
 as functions of $t$:  the nearest neighbor correlators $C_{\alpha \beta}$ and
 $C_{\alpha \gamma}$, and their difference.
 The difference $C_{\alpha \beta} - C_{\alpha \gamma}$ decreases from 1 as $t$ increases, and
 vanishes at the critical point where all sublattices are equally occupied. This
 gives an independent estimate of $t_c$, consistent with (\ref{eq_t_c}).

\begin{figure}[htb]
\centering
      \includegraphics[width=0.5\textwidth]{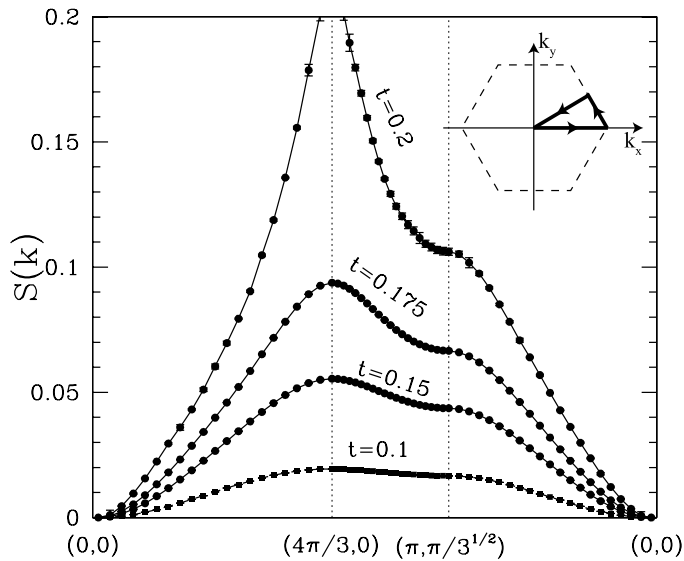}
      \caption{The static structure factor $S({\bf k})$ along high-symmetry cuts through the Brillouin
zone for various $t$ at $n=2/3$ filling. The inset shows the Brillouin zone, and path chosen. \label{fig_wei_k}}
\end{figure}

\begin{figure}[htb]
\centering
      \includegraphics[width=0.5\textwidth]{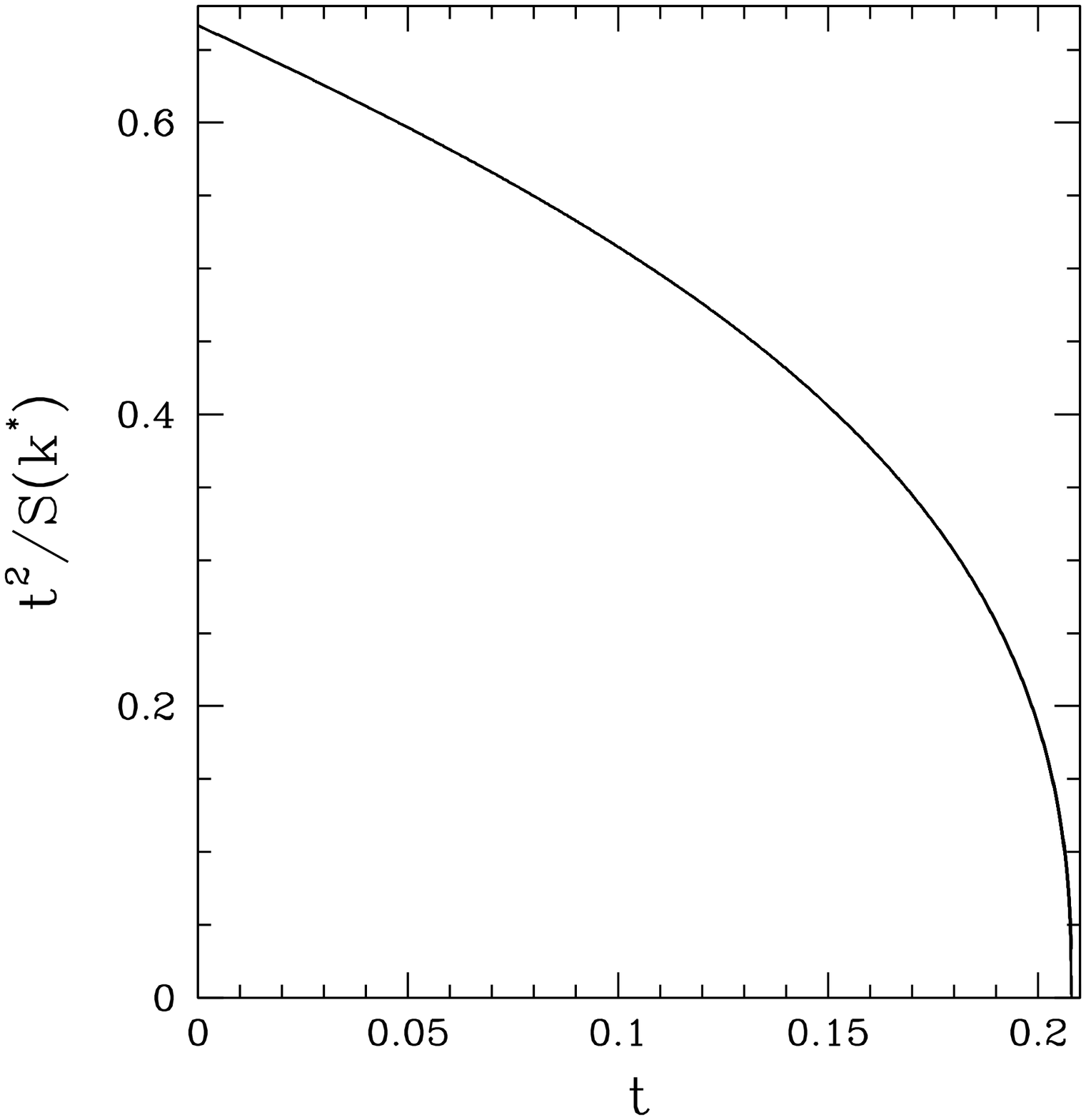}
      \caption{The static structure factor $S({\bf k}^*)$ versus $t$  at $n=2/3$ filling.
       Several different integrated differential approximants to
            the series  are shown.
      Since the series begins with a $t^2$ term we divide this out for convenience.
       \label{fig_S_max}}
\end{figure}

 Another quantity of interest is the static structure factor
 \be
 S({\bf k}) = {1\over N} \sum_{i,j} ( \langle n_i n_j \rangle -  \langle n_i \rangle \langle n_j \rangle )
 e^{i {\bf k} \cdot ( {\bf R}_{j} - {\bf R}_{i} ) }
 \ee
 This differs from the more usual definition (eq. Ref. \onlinecite{bon03}) by the subtraction of the second term, which
 is necessary to implement the linked cluster method efficiently. The effect of this
 term is to remove the $\delta$-function peak at ${\bf k}=0$,
 and reduce it at the special wavevector ${\bf k}^*$.
 At the transition point $t_c$, where $\langle n_i \rangle$ is independent of
 sublattice, the contribution of this second term vanishes, and our $S({\bf k})$
 is expected to show a $\delta$-function peak at ${\bf k}^*$.
 This is confirmed in Figure \ref{fig_wei_k}, where we show $S({\bf k})$
 curves along symmetry lines in the Brillouin zone for various $t$ values.

At ${\bf k}={\bf k}^*$, the Dlog Pad\'e approximants to the series $S({\bf k}^*)$ give a critical point
estimate $t_c=0.208(3)$ with exponent $-0.40(3)$ (see Table \ref{tab_pade}).
To obtain a  more accurate estimate of $S({\bf k}^*)$
versus $t$, we use  integrated differential approximants to the series
$[S({\bf k}^*)]^{-2.5}$ (which vanishes linearly at $t_c$). The result is shown in
Fig. \ref{fig_S_max}, from which we can locate the critical point more accurately, at
$t_c=0.208(1)$.

%
%
%

\section{The $n=\half$ Phase\label{sec4}}

As noted in Section \ref{sec2}, the case of half-filling ($n=\half$) is rather more subtle, as the
unperturbed system has an infinitely degenerate ground state. To allow a series expansion we
introduce an auxiliary field to break this degeneracy. There are many ways to do this, but
the simples choice is to decompose the triangular lattice into two sublattices ($\alpha , \beta$)
shown in Figure \ref{fig1}(c), and to impose a staggered field, which favours occupancy
of the $\alpha$ sublattice. Thus the unperturbed Hamiltonian is
\be
H_0 = V  \sum_{\langle ij\rangle } n_i n_j - h \sum_i \eta_i n_i
\ee
and the perturbation is
\be
H_1 = - t \sum_{\langle ij\rangle } ( a_i^\dagger a_j + a_j^\dagger a_i ) + h \sum_i \eta_i n_i
\ee
where $\eta_i =+1, -1$ on sublattices $\alpha$, $\beta$ respectively. The full Hamiltonian
is $H=H_0 + \lambda V$. Perturbation series are  derived in power of $\lambda$, and
$\lambda=1$ corresponds to the original Hamiltonian. Note that the field terms cancel in this limit.
The value of $h$ is arbitrary and can be chosen to obtain optimal convergence of the series.

\begin{figure}[htb]
\centering
      \includegraphics[width=0.5\textwidth]{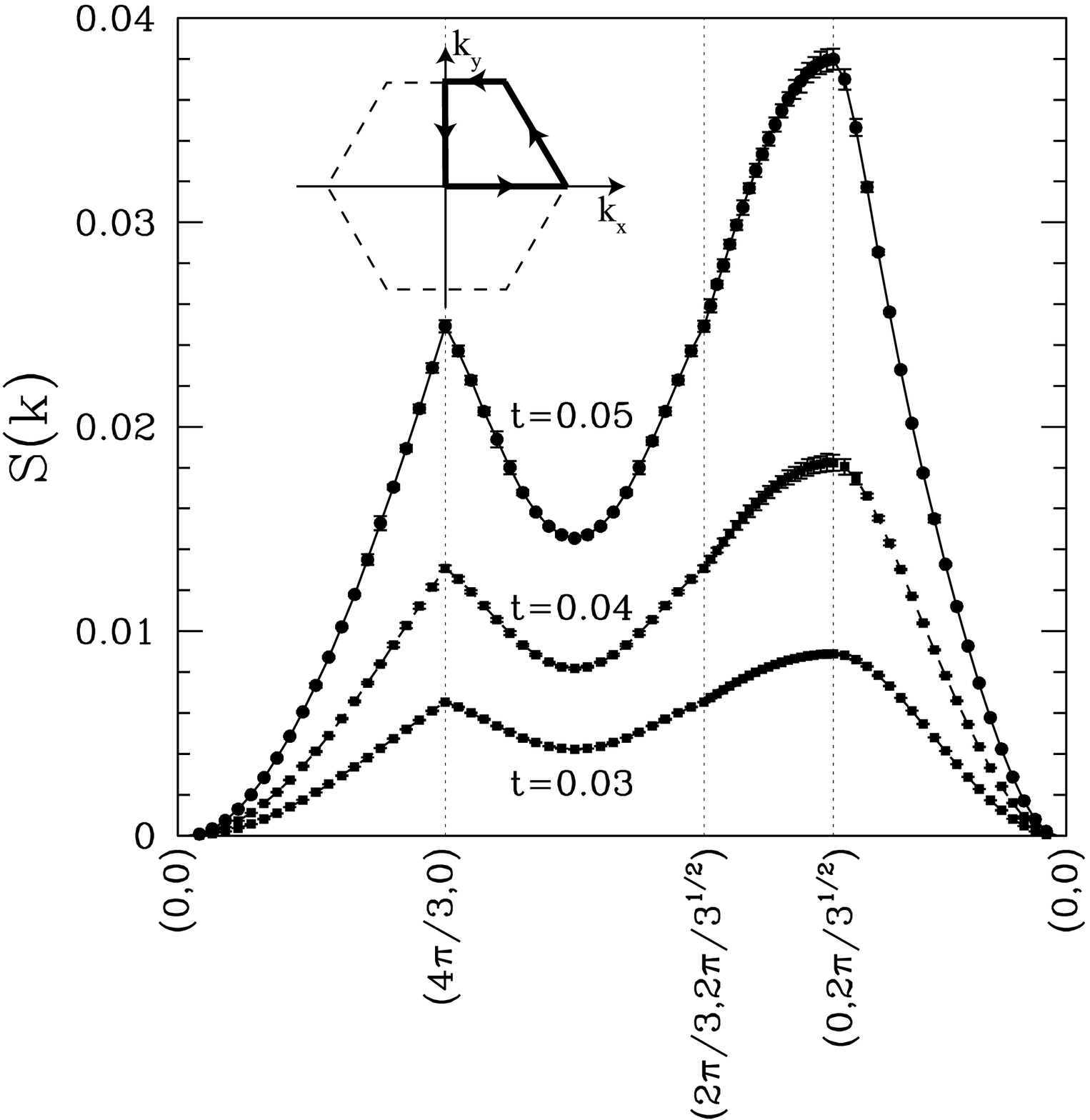}
      \caption{The static structure factor $S({\bf k})$ along high-symmetry cuts through the Brillouin
zone for $t=0.03, 0.04, 0.05$ at $n=1/2$ filling. The inset shows the Brillouin zone, and the path chosen.\label{fig_wei_k_hf}}
\end{figure}

We have derived series for the same quantities as in the previous Section: the ground state energy and
the order parameter, $m\equiv  \langle n_{\alpha} \rangle - \langle n_{\beta} \rangle$,
to order $\lambda^{11}$, pair correlations and the structure factor
to order $\lambda^{10}$. The series are evaluated as single variable series in $\lambda$, for
fixed values of $t$, $h$. To estimate the critical value $t_c$ we proceed as follows.
For any value of $t$ the series will have a singularity at $\lambda=\lambda_c (t)$
and $t_c$ is determined from the condition $\lambda_c(t_c)=1$.
For $t<t_c$, $\lambda_c$ is larger than 1, implying that the system retains the striped order shown in
 Figure \ref{fig1}(c) at $\lambda=1$, while for  $t>t_c$, $\lambda_c$ is less than 1.
The analysis
is less precise than for the $n=\frac{1}{3}$, $\frac{2}{3}$ cases, but
we obtain a consistent critical point at $t_c\simeq 0.06$.
Table \ref{tab_ser_hf} presents a number of series, for the parameter choice
$t=0.06$, $h=0.25$. Table  \ref{tab_pade_hf} presents the results of a Dlog Pad\'e approximant
analysis.
 The order parameter series, in particular, is quite well behaved and shows a consistent
sequence of estimates of $\lambda_c$ converging to $\lambda_c\simeq 1.0$.
The series for correlators and for the structure factor give a similar conclusion, that
$t_c\simeq 0.06$.

\begin{figure}[htb]
\centering
      \includegraphics[width=0.5\textwidth]{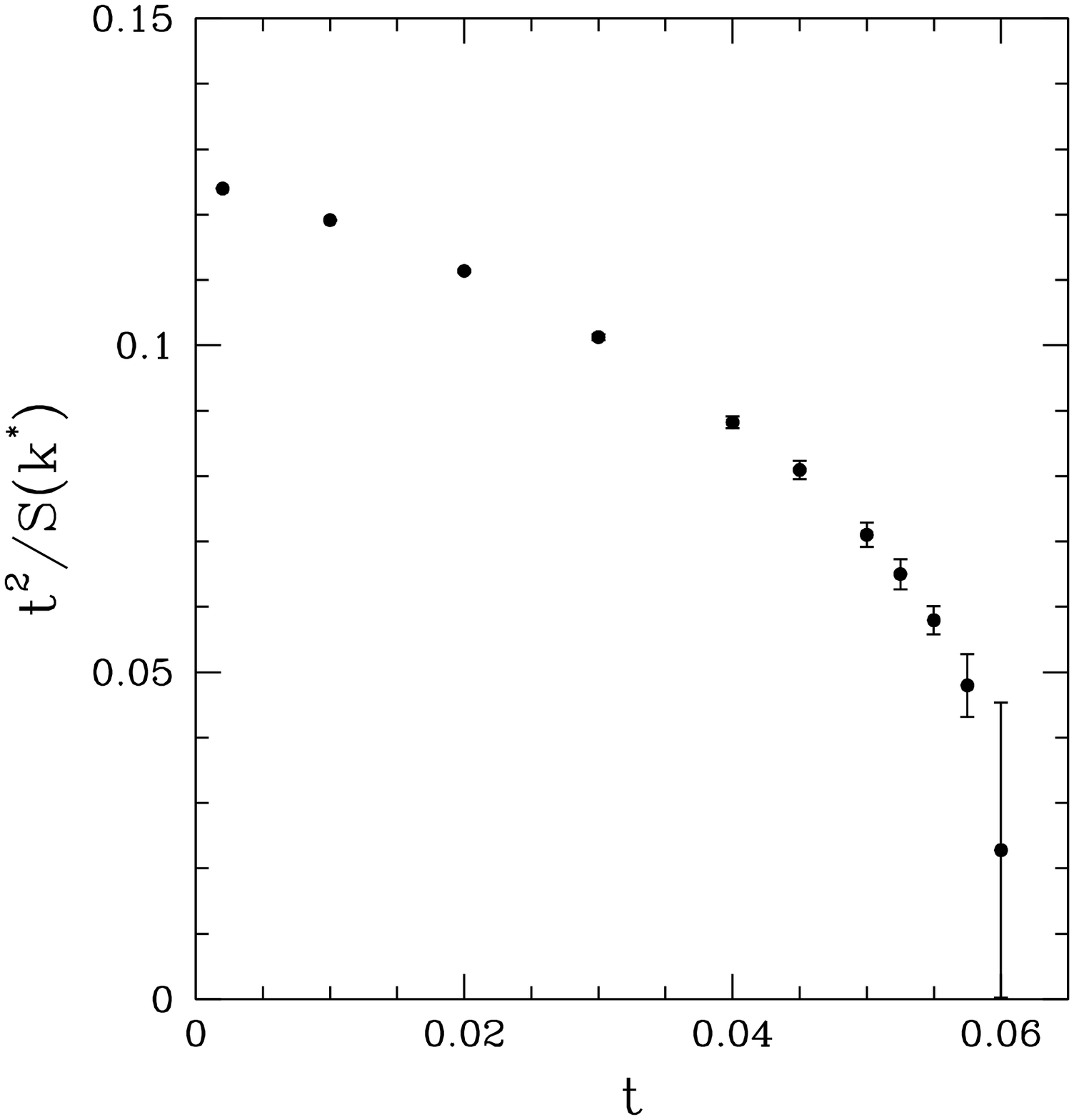}
      \caption{The static structure factor $S({\bf k}^*)$ versus $t$  at $n=1/2$ filling. \label{fig_S_max_hf}}
\end{figure}

In Figure \ref{fig_wei_k_hf} we show the structure factor along symmetry lines in the Brillouin zone for $n=\half$.
The ordering wavevector for the striped 2-sublattice ordered structure is
${\bf k}^* = 2 \pi (0, 1/\sqrt{3})$ and we see the development of a divergent peak at ${\bf k}^*$,
as $t\to 0.06$.
There is also a peak at the ordering wavevector for the $\sqrt{3}\times \sqrt{3}$ phase.
Figure \ref{fig_S_max_hf} shows the divergence of $S({\bf k}^*)$ as $t$ increases, towards 0.06.

Our estimate for the transition $t_c\sim 0.06$ is somewhat lower than the latest Monte Carlo
estimate\cite{wes05} of $t_c\sim 0.115$.
We discuss possible reasons for this in the final
section of the paper.


\section{Conclusions\label{sec5}}

We have used series expansions to investigate the ground state phase diagram of the hard-core
boson model on the triangular lattice, which is of considerable current interest.
Series are derived in powers of the hopping parameter $t$, and hence probe the spatially
ordered phases and transitions to a uniform superfluid phase.
Series have been computed for the ground state energy, the order parameters, the
near neighbour pair correlations and the static structure factor. All of
these quantities show consistent results  with each other.

At the commensurate densities $n=\frac{1}{3}, \frac{2}{3}$ we locate a quantum phase transition at
$(t/V)_c \simeq 0.208(1)$, in good agreement with the most recent Monte Carlo results of
$0.195\pm 0.025$. The order parameter appears to vanish at $(t/V)_c$ with a power low, with an exponent
$\beta \simeq 0.09$. At $n=\frac{1}{2}$ (half-filling) early Monte Carlo work\cite{bon03}
indicated no ordered solid phase at any finite $t$, although the most recent work\cite{wes05,hei05,mel05,bon05}
finds a supersolid phase persisting to $t/V\simeq 0.1$.
To derive series expansion for this case we stabilize one of the infinite manifold of ground states
(a `striped' 2-sublattice phase) by applying an external field. This phase, which presumably is supersolid,
appears stable to $t/V\sim 0.06$.

Using the present approach we are unable to probe off-diagonal long-range order and hence to study
the superfluid phases.

We are also unable to exclude the possibility of weakly first-order rather than second-order transitions.
In this context we mention recent field theoretical work\cite{bur05}, which argues that the transitions
in this model lie in the class of `deconfined quantum critical points'.

\begin{acknowledgments}
This work  forms part of a research project supported by
a grant from the Australian Research Council.
We are grateful for the computing resources provided
by the Australian Partnership for Advanced Computing (APAC)
National Facility and by the
Australian Centre for Advanced Computing and Communications (AC3).
\end{acknowledgments}

\newpage



\begin{table*}
\squeezetable
\caption{Series for ground state energy per site $E_0(t)$, order parameter $m$, difference of
shortest correlator $C_{\alpha \beta} - C_{\alpha\gamma}$, and static structure factor $S$ at ${\bf k}^*$
for the $n=2/3$ case.
Nonzero coefficients $t^n$ up to order $n=12$  are listed. }\label{tab_ser}
\begin{ruledtabular}
\begin{tabular}{|r|l|l|l|l|}
\multicolumn{1}{|c|}{$n$} & \multicolumn{1}{c|}{$E_0(t)$} & \multicolumn{1}{c|}{$m$}
& \multicolumn{1}{c|}{$C_{\alpha \beta} - C_{\alpha\gamma}$} & \multicolumn{1}{c|}{$S({\bf k}^*)$} \\
\hline
  0 &  ~~1.000000000                &  ~~1.000000000                &  ~~1.000000000                &  ~~0.000000000                \\
  1 &  ~~0.000000000                &  ~~0.000000000                &  ~~0.000000000                &  ~~0.000000000                \\
  2 & $-$1.000000000                & $-$2.250000000                & $-$2.750000000                &  ~~1.500000000                \\
  3 & $-$1.000000000                & $-$4.500000000                & $-$5.500000000                &  ~~3.000000000                \\
  4 & $-$9.500000000$\times 10^{-1}$& $-$1.019750000$\times 10^{1}$ & $-$1.052083333$\times 10^{1}$ &  ~~8.168333333                \\
  5 & $-$2.355555556                & $-$3.750666667$\times 10^{1}$ & $-$3.819305556$\times 10^{1}$ &  ~~3.518166667$\times 10^{1}$ \\
  6 & $-$7.937379630                & $-$1.418886375$\times 10^{2}$ & $-$1.498124375$\times 10^{2}$ &  ~~1.428442162$\times 10^{2}$ \\
  7 & $-$2.616389136$\times 10^{1}$ & $-$5.619014174$\times 10^{2}$ & $-$5.933175977$\times 10^{2}$ &  ~~6.107654281$\times 10^{2}$ \\
  8 & $-$9.005836612$\times 10^{1}$ & $-$2.307990196$\times 10^{3}$ & $-$2.428296319$\times 10^{3}$ &  ~~2.667494106$\times 10^{3}$ \\
  9 & $-$3.229654053$\times 10^{2}$ & $-$9.596323715$\times 10^{3}$ & $-$1.008214369$\times 10^{4}$ &  ~~1.170747249$\times 10^{4}$ \\
 10 & $-$1.201486606$\times 10^{3}$ & $-$4.067064188$\times 10^{4}$ & $-$4.264450526$\times 10^{4}$ &  ~~5.234712003$\times 10^{4}$ \\
 11 & $-$4.607510926$\times 10^{3}$ & $-$1.750821492$\times 10^{5}$ & $-$1.832564264$\times 10^{5}$ &                 \\
 12 & $-$1.792870548$\times 10^{4}$ & $-$7.572580451$\times 10^{5}$ &                    &                  \\
\end{tabular}
\end{ruledtabular}
\end{table*}

\begin{table*}
\squeezetable
\caption{
$[n/m]$ Dlog Pad\'e approximants to the series of
$d^2 E_0/d t^2$, $m$, $C_{\alpha \beta} - C_{\alpha\gamma}$, and $S({\bf k}^*)$ for $n=2/3$ case.
The position of the
singularity (pole), the critical index (index) from unbiased approximants are given.
An asterisk denotes a defective approximant.
 }\label{tab_pade}
\begin{ruledtabular}
\begin{tabular}{rlllll}
\multicolumn{1}{c}{n} &\multicolumn{1}{c}{$[(n-2)/n]$}&\multicolumn{1}{c}{$[(n-1)/n]$}
&\multicolumn{1}{c}{$[n/n]$} &\multicolumn{1}{c}{$[(n+1)/n]$}&\multicolumn{1}{c}{$[(n+2)/n]$}
 \\
\multicolumn{1}{c}{} &\multicolumn{1}{c}{pole (index)}
&\multicolumn{1}{c}{pole (index)}
&\multicolumn{1}{c}{pole (index)}
&\multicolumn{1}{c}{pole (index)}
&\multicolumn{1}{c}{pole (index)} \\
\hline
\multicolumn{6}{c}{$d^2 E_0(t)/dt^2$ }\\
n=1 &                & 1.2500($-$3.750)&  0.0518($-$0.006)  & 0.1834($-$0.286) &  0.1929($-$0.350)  \\
n=2 & 0.2342($-$0.388) & 0.1514($-$0.139)&  0.1936($-$0.357)  & 0.1953($-$0.373) &  0.1921($-$0.345)* \\
n=3 & 0.1726($-$0.213) & 0.1951($-$0.371)&  0.1933($-$0.355)* & 0.2124($-$0.845) &  0.2007($-$0.438)  \\
n=4 & 0.2111($-$0.691) & 0.2020($-$0.463)&  0.2010($-$0.444)  & 0.2036($-$0.502)   \\
n=5 & 0.2006($-$0.438) & 0.2018($-$0.459) \\
\hline
\multicolumn{6}{c}{$m$ }\\
n=1 &                &                & 0.3333(0.5000) &  0.2651(0.2517) & 0.2138(0.1064) \\
n=2 &                & 0.2750(0.2896) & 0.1673(0.0261) &  0.2209(0.1246) & 0.2180(0.1162) \\
n=3 & 0.2368(0.1741) & 0.2257(0.1411) & 0.2164(0.1109) &  0.1999(0.0504) & 0.2126(0.0986) \\
n=4 & 0.1852(0.0262) & 0.2103(0.0897) & 0.2116(0.0946) &  0.2108(0.0910) & 0.2087(0.0809) \\
n=5 & 0.2121(0.0965) & 0.2111(0.0925) & 0.2168(0.0994)*&  0.2100(0.0876) \\
n=6 & 0.2096(0.0860) & 0.2104(0.0895) \\
\hline
\multicolumn{6}{c}{$C_{\alpha \beta} - C_{\alpha\gamma}$ }\\
n=1 &                &                & 0.3333(0.6111)  & 0.2884(0.3959) & 0.2146(0.1213) \\
n=2 &                & 0.2932(0.4219) & 0.1049(0.0019)  & 0.2207(0.1393) & 0.2182(0.1309) \\
n=3 & 0.2431(0.2229) & 0.2288(0.1712) & 0.2166(0.1247)  & 0.1983(0.0487) & 0.2134(0.1134) \\
n=4 & 0.1906(0.0404) & 0.2115(0.1047) & 0.2123(0.1079)  & 0.2111(0.1025) & 0.2090(0.0905) \\
n=5 & 0.2124(0.1087) & 0.2118(0.1058) & 0.1988(0.0307)* \\
n=6 & 0.2100(0.0970) \\
\hline
\multicolumn{6}{c}{$S({\bf k}^*)$ }\\
n=1 &                  & 0.2902($-$0.5805) & 0.1508($-$0.1568) & 0.2228($-$0.5051) & 0.2001($-$0.3291) \\
n=2 &  0.1833($-$0.2678) & 0.1941($-$0.3066) & 0.2052($-$0.3709) & 0.2051($-$0.3700) & 0.2152($-$0.5228) \\
n=3 & *                & 0.2051($-$0.3700) & 0.2052($-$0.3709) & 0.2072($-$0.3895) \\
n=4 &  0.2082($-$0.4033) & 0.2081($-$0.4024)  \\
\end{tabular}
\end{ruledtabular}
\end{table*}

\begin{table*}
\squeezetable
\caption{Series for ground state energy per site $E_0$, order parameter $m$, difference of
shortest correlator $C_{\alpha \alpha} - C_{\alpha\beta}$ and $C_{\alpha \alpha} - C_{\beta\beta}$,
 and static structure factor $S$ at ${\bf k}^*=(0,2\pi/\sqrt{3})$ for
$t=0.06$ and the field $h=0.25$ at $n=1/2$ case.
Nonzero coefficients $\lambda^n$ up to order $n=11$  are listed. }\label{tab_ser_hf}
\begin{ruledtabular}
\begin{tabular}{|r|l|l|l|l|l|}
\multicolumn{1}{|c|}{$n$} & \multicolumn{1}{c|}{$E_0$} & \multicolumn{1}{c|}{$m$}
& \multicolumn{1}{c|}{$C_{\alpha \alpha} - C_{\alpha\beta}$}
& \multicolumn{1}{c|}{$C_{\alpha \alpha} - C_{\beta\beta}$}
& \multicolumn{1}{c|}{$S({\bf k}^*)$} \\
\hline
  0 &  ~~5.000000000$\times 10^{-1}$ &  ~~1.000000000                 &  ~~1.000000000                 &  ~~1.000000000                 &  ~~0.000000000                 \\
  1 &  ~~0.000000000                 &  ~~0.000000000                 &  ~~0.000000000                 &  ~~0.000000000                 &  ~~0.000000000                 \\
  2 & $-$4.800000000$\times 10^{-3}$ & $-$1.280000000$\times 10^{-2}$ & $-$1.760000000$\times 10^{-2}$ & $-$1.280000000$\times 10^{-2}$ &  ~~1.280000000$\times 10^{-2}$ \\
  3 & $-$1.984000000$\times 10^{-3}$ & $-$1.058133333$\times 10^{-2}$ & $-$1.454933333$\times 10^{-2}$ & $-$1.058133333$\times 10^{-2}$ &  ~~1.058133333$\times 10^{-2}$ \\
  4 & $-$8.346453333$\times 10^{-4}$ & $-$6.818781867$\times 10^{-3}$ & $-$9.282071467$\times 10^{-3}$ & $-$6.818781867$\times 10^{-3}$ &  ~~6.969821867$\times 10^{-3}$ \\
  5 & $-$3.763032178$\times 10^{-4}$ & $-$4.322561214$\times 10^{-3}$ & $-$5.768006277$\times 10^{-3}$ & $-$4.322561214$\times 10^{-3}$ &  ~~4.758822760$\times 10^{-3}$ \\
  6 & $-$1.920603977$\times 10^{-4}$ & $-$2.973745768$\times 10^{-3}$ & $-$3.879098112$\times 10^{-3}$ & $-$2.973745768$\times 10^{-3}$ &  ~~3.673001969$\times 10^{-3}$ \\
  7 & $-$1.129137214$\times 10^{-4}$ & $-$2.252261754$\times 10^{-3}$ & $-$2.888143560$\times 10^{-3}$ & $-$2.252261754$\times 10^{-3}$ &  ~~3.111869445$\times 10^{-3}$ \\
  8 & $-$7.425851217$\times 10^{-5}$ & $-$1.819385962$\times 10^{-3}$ & $-$2.309557263$\times 10^{-3}$ & $-$1.819385962$\times 10^{-3}$ &  ~~2.744159087$\times 10^{-3}$ \\
  9 & $-$5.242495609$\times 10^{-5}$ & $-$1.520557155$\times 10^{-3}$ & $-$1.918242621$\times 10^{-3}$ & $-$1.520557155$\times 10^{-3}$ &  ~~2.454367694$\times 10^{-3}$ \\
 10 & $-$3.867629129$\times 10^{-5}$ & $-$1.296181354$\times 10^{-3}$ & $-$1.627204515$\times 10^{-3}$ & $-$1.296181354$\times 10^{-3}$ &  ~~2.215862063$\times 10^{-3}$ \\
 11 & $-$2.946020021$\times 10^{-5}$ & $-$1.122501973$\times 10^{-3}$ &                  &                  &                   \\
\end{tabular}
\end{ruledtabular}
\end{table*}

\begin{table*}
\squeezetable
\caption{
$[n/m]$ Dlog Pad\'e approximants to the series of
$d^2 E_0/d \lambda^2$, $m$, $C_{\alpha \alpha} - C_{\alpha\beta}$, $C_{\alpha \alpha} - C_{\beta\beta}$, and $S({\bf k}^*)$ for
$t=0.06$ and the field $h=0.25$ at $n=1/2$ case.
The position of the
singularity (pole), the critical index (index) from unbiased approximants are given.
An asterisk denotes a defective approximant.
 }\label{tab_pade_hf}
\begin{ruledtabular}
\begin{tabular}{rlllll}
\multicolumn{1}{c}{n} &\multicolumn{1}{c}{$[(n-2)/n]$}&\multicolumn{1}{c}{$[(n-1)/n]$}
&\multicolumn{1}{c}{$[n/n]$} &\multicolumn{1}{c}{$[(n+1)/n]$}&\multicolumn{1}{c}{$[(n+2)/n]$}
 \\
\multicolumn{1}{c}{} &\multicolumn{1}{c}{pole (index)}
&\multicolumn{1}{c}{pole (index)}
&\multicolumn{1}{c}{pole (index)}
&\multicolumn{1}{c}{pole (index)}
&\multicolumn{1}{c}{pole (index)} \\
\hline
\multicolumn{6}{c}{$d^2 E_0/d\lambda^2$ }\\
n=1 &               & 2.2586($-$2.8007) & 1.4547($-$1.1617) & 0.9731($-$0.3478) & 0.9220($-$0.2803) \\
n=2 & 1.6180($-$1.5630) & 0.7556($-$0.1148) & 0.9164($-$0.2716) & 0.9458($-$0.3145) & 1.1581($-$6.1720)* \\
n=3 & 0.9303($-$0.2941) & 1.0207($-$0.4982) & 1.0060($-$0.4512) & 0.9990($-$0.4284) & 0.9929($-$0.4076) \\
n=4 & 1.0063($-$0.4522) & 0.9889($-$0.3895) & 0.9707($-$0.3098) &               &               \\
n=5 & 0.9711($-$0.3113) &               &               &               &               \\
\hline
\multicolumn{6}{c}{$m$ }\\
n=1 &               & * & 0.8065(0.0166) & 1.1500(0.0483) & 1.2383(0.0649) \\
n=2 &              *& 1.3497($-$0.0279)*& 1.2735(0.0755) & 1.2075(0.0577) & 1.3029(0.0748)*\\
n=3 & 1.1914(0.0521) & 1.1288(0.0392) & 1.0916(0.0316) & 1.0711(0.0271) & 1.0576(0.0241) \\
n=4 & 1.0746(0.0277) & 1.0591(0.0243) & 1.0364(0.0187) & 1.0094(0.0124) & 0.9977(0.0100) \\
n=5 & 0.9887(0.0083) & 0.9929(0.0090) & 0.9946(0.0094) &               &               \\
n=6 & 0.9962(0.0097) &               &               &               &               \\
\hline
\multicolumn{6}{c}{$C_{\alpha \alpha} - C_{\alpha\beta}$ }\\
n=1 &               & * & 0.8065(0.0229) & 1.1563(0.0675) & 1.2532(0.0931) \\
n=2 &              *& 1.3327($-$0.0378)*& 1.2969(0.1123) & 1.2229(0.0830) & 1.3011(0.1043)*\\
n=3 & 1.2075(0.0751) & 1.1419(0.0559) & 1.1015(0.0443) & 1.0783(0.0373) & 1.0631(0.0326) \\
n=4 & 1.0820(0.0382) & 1.0643(0.0327) & 1.0407(0.0251) & 1.0149(0.0171) &               \\
n=5 & 1.0014(0.0133) & 0.9989(0.0127) &               &               &               \\
\hline
\multicolumn{6}{c}{$C_{\alpha \alpha} - C_{\beta\beta}$  }\\
n=1 &               & * & 0.8065(0.0166) & 1.1500(0.0483) & 1.2383(0.0649) \\
n=2 &              *& 1.3497($-$0.0279)*& 1.2735(0.0755) & 1.2075(0.0577) & 1.3029(0.0749)*\\
n=3 & 1.1914(0.0521) & 1.1288(0.0392) & 1.0916(0.0316) & 1.0711(0.0271) & 1.0576(0.0241) \\
n=4 & 1.0746(0.0277) & 1.0591(0.0242) & 1.0364(0.0187) & 1.0094(0.0124) &               \\
n=5 & 0.9887(0.0083) & 0.9929(0.0090) &               &               &               \\
\hline
\multicolumn{6}{c}{$S({\bf k}^*)$ }\\
n=1 &               & 2.0378($-$1.6846) & 1.2297($-$0.6135) & 0.9509($-$0.2837) & 0.9605($-$0.2952) \\
n=2 & 1.4031($-$0.8844) & 0.8620($-$0.1860) & 0.9602($-$0.2948) & 0.9482($-$0.2815)*& 1.0447($-$0.4819) \\
n=3 & 0.9726($-$0.3144) & 1.0375($-$0.4476) & 1.0246($-$0.4121) & 1.0163($-$0.3890) &               \\
n=4 & 1.0250($-$0.4132) & 0.9918($-$0.3047) &               &               &               \\
\end{tabular}
\end{ruledtabular}
\end{table*}

\end{document}